\documentclass[a4paper,
               keeplastbox,   
               ]{jacow}
\usepackage{pdfpages,multirow,ragged2e,siunitx} %
\makeatletter%
	\ifboolexpr{bool{xetex}}
	 {\renewcommand{\Gin@extensions}{.pdf,%
	                    .png,.jpg,.bmp,.pict,.tif,.psd,.mac,.sga,.tga,.gif,%
	                    .eps,.ps,%
	                    }}{}
\makeatother

\ifboolexpr{bool{xetex} or bool{luatex}} 
 {}                                      
 {\usepackage[utf8]{inputenc}}           

\usepackage[USenglish]{babel}

\usepackage{subfig}

\ifboolexpr{bool{jacowbiblatex}}%
 {%
  \addbibresource{jacow-test.bib}
  \addbibresource{biblatex-examples.bib}
 }{}
\begin{document}

\title{Design concept for the second interaction region for Electron-Ion Collider\thanks{This material is based upon work supported by Jefferson Science Associates, LLC under Contract No. DE-AC05-06OR23177 and Brookhaven Science Associates, LLC under Contract No. DE-SC0012704 with the U.S. Department of Energy.}}

\author{B.R.\ Gamage\thanks{randika@jlab.org},  V.\ Burkert, R.\ Ent, Y.\ Furletova, D.\ Higinbotham, A.\ Hutton, F.\ Lin,\\ T.\ Michalski,
V.S.\ Morozov,  R.\ Rajput-Ghoshal, D.\ Romanov,   T.\ Satogata, A.\ Seryi, A.\ Sy, \\ C.\ Weiss, M.\ Wiseman, W.\ Wittmer, Y.\ Zhang, 
Jefferson Lab, Newport News, VA, USA \\
E.-C.\ Aschenauer, J.S.\ Berg, A.\ Jentsch, A.\ Kiselev,  C.\ Montag, R.\ Palmer, B.\ Parker, \\ V.\ Ptitsyn, F.\ Willeke, H.\ Witte, 
Brookhaven National Lab, Upton, NY, USA \\
C.\ Hyde, Old Dominion University, Norfolk, VA, USA \\
P.\ Nadel-Turonski, Stony Brook University, Stony Brook, NY, USA
}

\maketitle

\begin{abstract}
The possibility of two interaction regions (IRs) is a design requirement for the Electron Ion Collider (the EIC)~\cite{longrange}. There is also a significant interest from the nuclear physics community in a 2nd IR with measurements capabilities complementary to those of the first IR. While the 2nd IR will be in operation over the entire energy range of \SI{\sim 20}{GeV} to \SI{\sim 140}{GeV} center of mass (CM). The 2nd IR can also provide an acceptance coverage complementary to that of the first. We present a brief overview and the current progress of the 2nd IR design in terms of the parameters, magnet layout, and beam dynamics. 
\end{abstract}
\section{Introduction}
A second EIC IR that is complementary to the performance and detection capabilities of the first IR has gained significant interest from the nuclear physics community. ``Complementary'' in this context means that both the energy range luminosity optimization and detection capabilities are complementary to the physics reach of the first IR. The main parameters critical to both detector and machine aspects of the design and their comparison to the first IR is shown in Table~\ref{table:summary}.
\subsection{Detection Requirements}
\begin{figure}[t]
   \centering
   \includegraphics*[width=55mm]{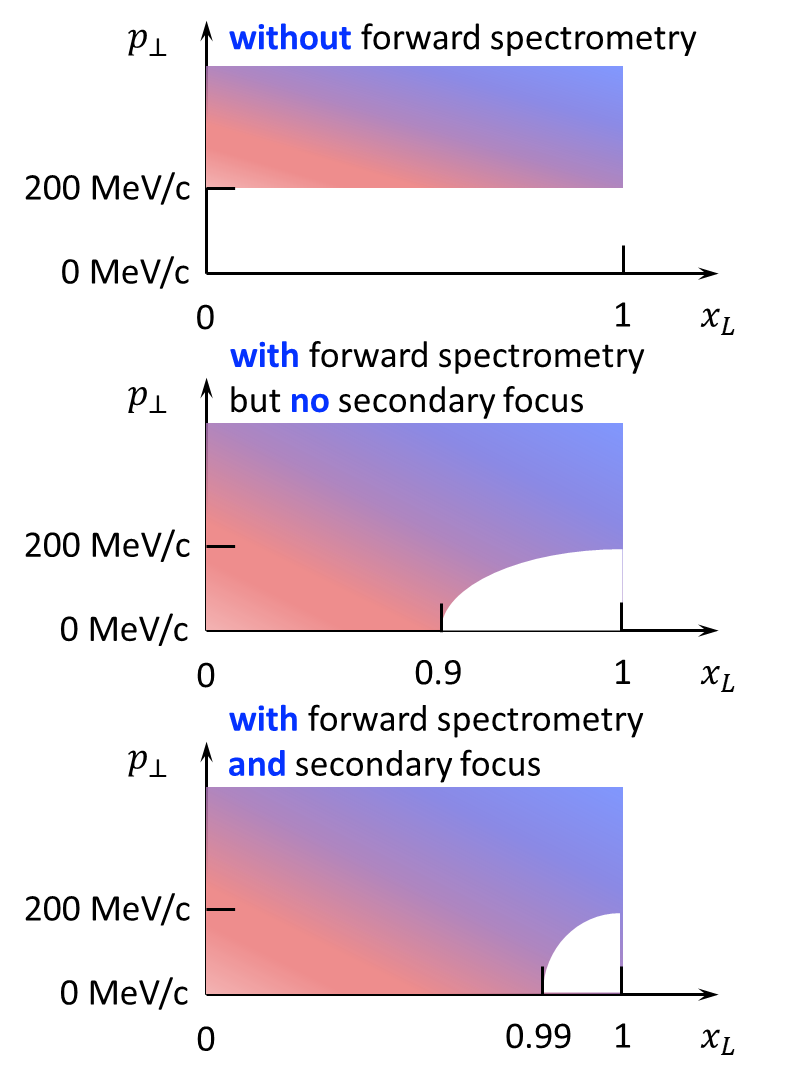}
   \caption{Illustration of the impact of the forward spectrometry and secondary focus on detector acceptance in the $\textit{x}_L - \textit{p}_T$ space for \SI{275}{\giga\electronvolt} protons.}
   \label{fig:acc}
\end{figure}

The 2nd IR includes a forward spectrometer with a secondary focus in x and y at a location with a large, flat dispersion, making it possible to detect particles with very small changes in rigidity~\cite{Morozov:IPAC2019-WEPGW123}. This includes ions that have lost only one neutron, and protons with a small longitudinal momentum loss (large $x_L$) even when their $p_T$ is small, where $x_L = p'_L / p_{\rm beam}$ and $p_T$ is the hadron transverse momentum~\cite{xlpt}. The zero-degree calorimeter (ZDC) acceptance for neutrons is \SI{\pm 7}{\milli\radian}. Protons with scattering angles of \SI{\pm 5}{\milli\radian} or less are detected downstream of the quadrupoles, while those with larger angles are detected in the B0 dipole in-between the central detector and the quadrupoles. The impact of having both forward spectrometry and secondary focus on the detector acceptance in the $x_L-p_T$ space is shown in Fig.~\ref{fig:acc}.
\begin{table*}[!hbt]
   \centering\small
   \caption{Summary of 2nd IR Design Requirements and Their Comparison to the 1st IR~\cite{1stIR}}
    \label{table:summary}
   \begin{tabular}{p{0.36\linewidth} | p{0.09\linewidth} | p{0.16\linewidth} |p{0.29\linewidth}}
       \toprule
       \textbf{Parameter} & \textbf{EIC IR 1} & \textbf{EIC IR 2} & \textbf{Impact} \\
         \midrule
        Energy Range   &	   &	   &	   \\
        ~~~~electrons [GeV]   &	5-18   &	5-18   &	\multirow{2}{*}{Facility operation}  \\
        ~~~~protons[GeV]   &	41,~100-275   &	41,~100-275   &	 \\
        Crossing angle [mrad]   &	25   &	35   &	$\textit{p}_T$ resolution, acceptance, geometry.   \\
        Detector space symmetry [m]   &	-4.5/+5   &	-5.5/+5.5   &	Forward/rear acceptance balance.   \\
        Forward angular acceptance [mrad]   &	20   &	25   &	Spectrometer dipole aperture.	   \\
        Far-forward angular acceptance [mrad]   &	4.5   &	5,7   &	Neuton cone, $\textit{p}_T^{max}$. \\
        Minimum $\Delta(B\rho)$/$(B\rho)$ allowing for detection of $\textit{p}_T = 0$ fragments & 0.1 & 0.003-0.01 & Beam focus with dispersion,\newline reach in $\textit{x}_L$ and $\textit{p}_T$ resolution, reach in $\textit{x}_B$ for exclusive process. \\
        Angular beam divergence at IP, h/v,rms [mrad] & 0.1/0.2 & < 0.2 & $\textit{p}_T^{min}$, $\textit{p}_T$ resolution. \\
        Low $Q^2$ electron acceptance & < 0.1 & < 0.1 & Not a hard requirement. \\
       \midrule
       \bottomrule
   \end{tabular}
\end{table*}
\subsection{Machine Requirements}

From a lattice design and beam dynamics points of view, the second IR must satisfy the following constraints: 1)~The IR beam lines are able to at least transport electron and ion beams over the entire energy ranges; 2)~The aperture-edge fields of all magnets are below \SI{4.6}{\tesla}; 3)~The IR optics is matched to the regular ring lattice. Modification of the present RHIC straight section can be done up to \SI{\sim 170}{\meter} from the collision point; 4)~With the 2nd IR included, the dynamic apertures and momentum acceptances of the electron and ion collider rings must be sufficiently large;  5)~The dispersion at the crab cavities satisfies $H_x \leq 15$\si{m} in the proton ring and $H_x \leq 2$\si{m} in the electron ring as established by beam-beam studies, where $H_x$ is the horizontal dispersion invariant; 6)~Beam aperture is greater than $10\sigma$ of the beam size for protons, greater than $15\sigma_x$ and $20\sigma_y$ for electrons; 7)~The synchrotron radiation generated by the electron beam does not affect the operation of the IR components; 8)~Does not significantly influence the beam life time of the machine; 9)~Re-use as many RHIC magnets as possible for matching into the arcs.
\begin{figure*}[!htb]
   \centering
   \includegraphics*[width=140mm]{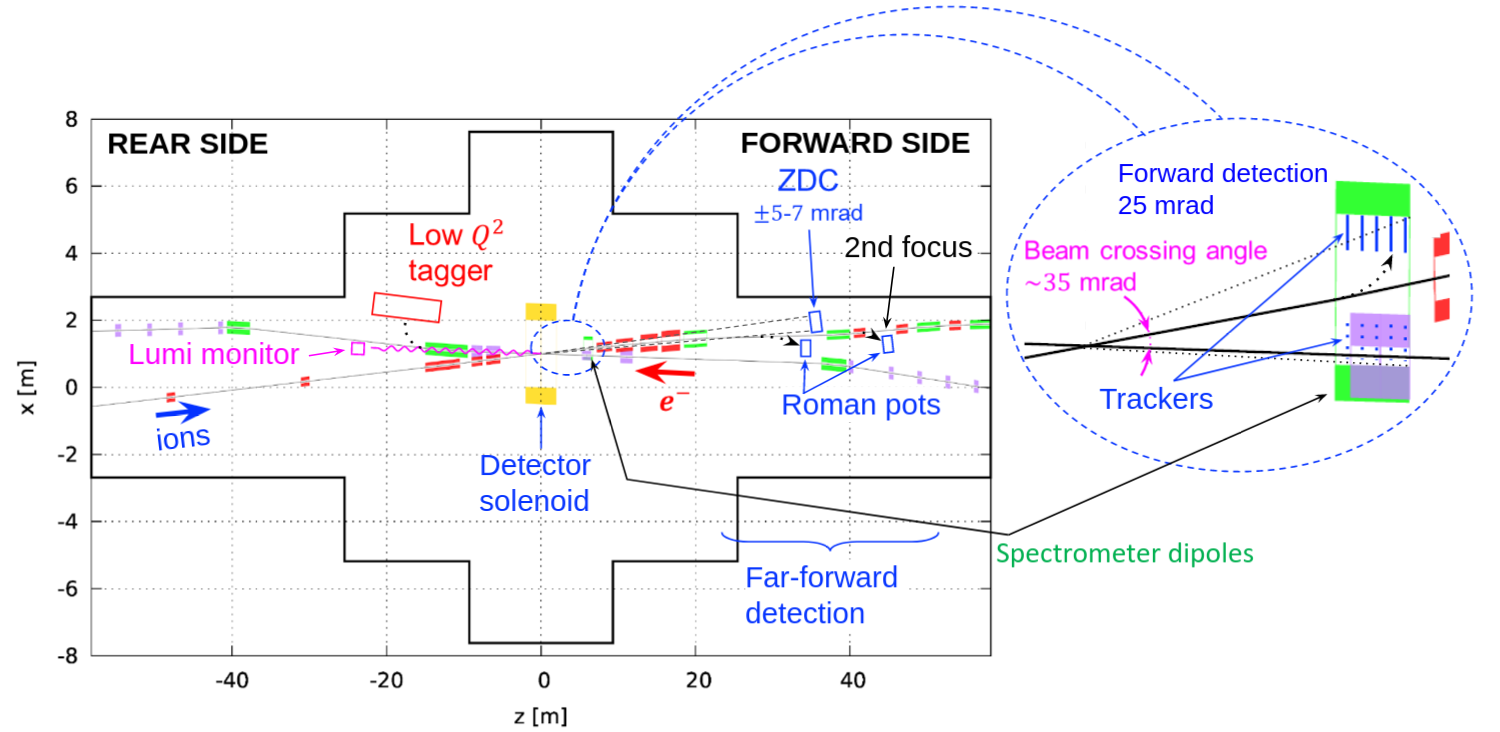}
   \caption{Layout of the second IR with a \SI{35}{\milli\radian} crossing angle indicating locations of the main forward and auxiliary detector component. The black solid lines outlines the size of the IP8 experimental hall and adjacent tunnel.}
   \label{fig:layout}
\end{figure*}
\subsection{Geometric Constraints}
The 2nd IR design must fit in the existing experimental hall geometry and accommodate detector components including a ZDC (60$\times$60$\times$200~\si{cm^3}), Roman pots, a low $Q^2$ tagger, off-momentum detectors and a luminosity monitor. At the far ends, space should be reserved to accommodate two spin rotators and one snake (each requires \SI{\sim 11}{\meter}). The IP needs to be shifted \SI{\sim1}{\meter} towards the inner wall to accommodate a bypass of the rapid cycling synchrotron (RCS) electron injector. These considerably constrain the IR crossing angle.
\section{IR layout}
The crossing angle of \SI{35}{\milli\radian} was selected based on constraints mentioned above and other factors. A larger crossing angle would require additional crab cavities which leads to additional cost, impedances, and other beam dynamic issues. A smaller crossing angle causes the beam line separation to be too low to place the required detectors without interfering with the electron beamline. Placing the ZDC detector between the two beamlines would require the hadron beamline to bend towards the inner wall of the experimental hall. The remaining space is not sufficient for this layout to be a feasible design. Hence the optimum crossing angle was chosen to be \SI{35}{\milli\radian} and the corresponding layout is shown in Fig.~\ref{fig:layout}.
\section{Linear Optics}
\subsection{Hadron Beam Optics}
\begin{figure}[!htb]
   \centering
   \includegraphics*[width=\columnwidth]{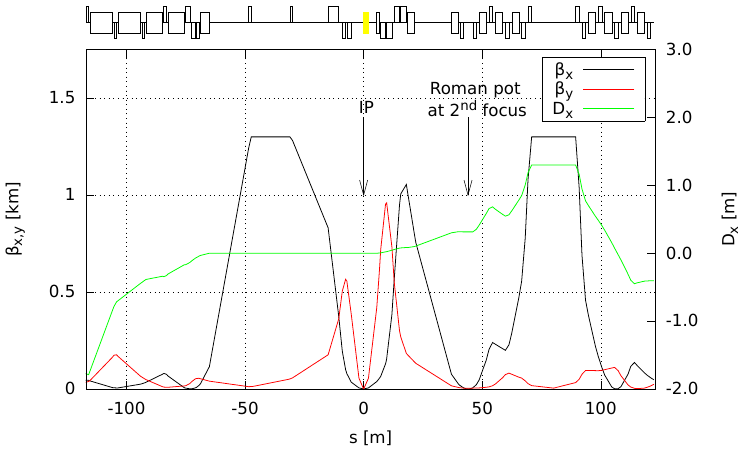}
   \caption{Twiss parameters for \SI{275}{\giga\electronvolt} protons. Horizontal and vertical $\beta$ functions (left vertical scale) and horizontal dispersion $D_x$ (right vertical scale) are plotted versus the distance along the beam $s$. Positions of the IP and of the Roman Pot at the secondary focus are indicated.}
   \label{fig:h_optics}
\end{figure}

The optics of the 2nd IR must support optimization of the luminosity and detection in different collider configurations, including different beam energies, ion species, and detector solenoid strengths. Fig.~\ref{fig:h_optics} shows a concept of the ion IR optics designed to support these and other detection and machine requirements described above. There are three physical quads upstream and four physical quads downstream of the IP. Their lengths and apertures have been optimized so that the 2nd IR can operate to the top energy of \SI{275}{\giga\electronvolt} while not exceeding an aperture edge field of \SI{4.6}{\tesla}. The first \SI{-6}{\milli\radian} spectrometer dipole is in front of the downstream final focusing block (FFB). It is used for momentum analysis of the forward scattered particles in the low-momentum-resolution region of the solenoid near its axis. Two additional spectrometer dipoles are located after the downstream FFB. They momentum-analyze small-momentum-offset particles that are moving very close to or within the ion beam. The dipoles are followed by a 15 m long machine-element-free straight section instrumented with several stages of Roman pots and off-momentum detectors. The area of the secondary focus is designed with \SI{\sim 0.5}{\meter} zero-slope dispersion and a strong beam focus which allows for the placement of Roman Pot detectors close to the beam with good $p_T$ resolution. 
\subsection{Electron Beam Optics}
\begin{figure}[!htb]
   \centering
   \includegraphics*[width=\columnwidth]{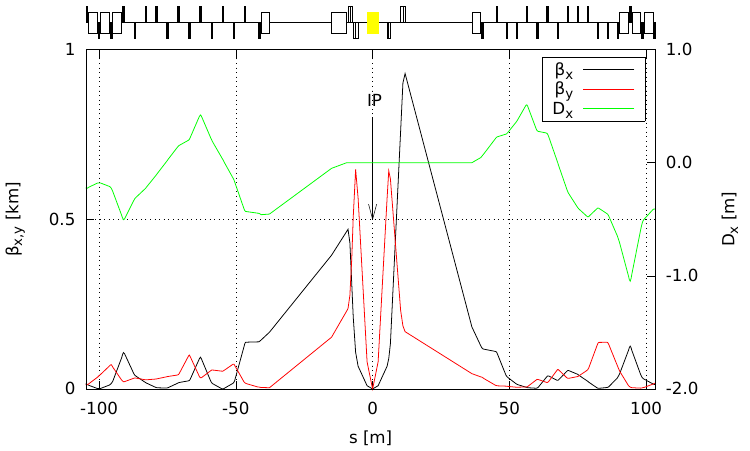}
   \caption{Twiss parameters for \SI{18}{\giga\electronvolt} electrons. Horizontal and vertical $\beta$ functions (left vertical scale) and horizontal dispersion $D_x$ (right vertical scale) are plotted versus the distance along the beam $s$. Position of the IP is indicated.}
   \label{fig:e_optics}
\end{figure}

The electron beamline in the second IR is similar to the first IR design. Special attention is given to preserve the geometric constraints imposed by the electron spin rotators and IP. Fig.~\ref{fig:e_optics} shows the current design for the electron optics at the 2nd IR. The rear side of the electron IR has a luminosity monitor and a low-$Q^2$ tagger. Low-$Q^2$ tagging is done using a spectrometer dipole placed on the rear side right after the electron FFQs.
\section{Nonlinear Dynamics}
\begin{figure}[!htb]
   \centering
   \includegraphics*[width=\columnwidth]{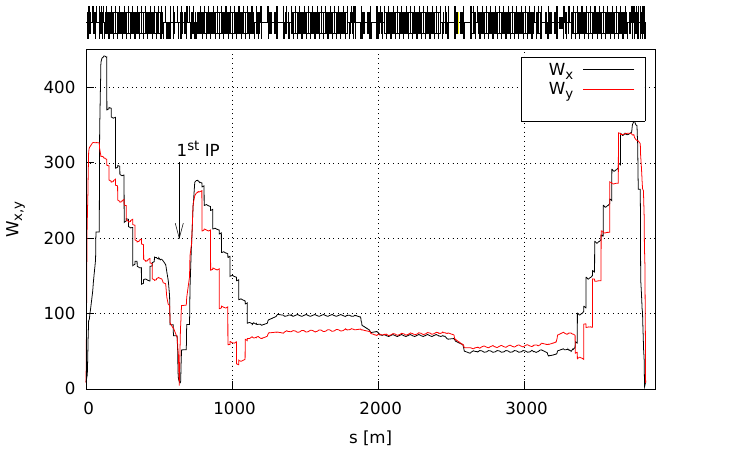}
   \caption{A preliminary result of the chromaticity correction with two IRs in the hadron ring for \SI{275}{\giga\electronvolt} protons. $2^{nd}$ IP is located at $s=0$. Where W-function $\sim\delta\beta/(\beta\delta)$.}
   \label{fig:chrom}
\end{figure}

The chromaticities generated by the IRs require dedicated correction schemes. If not locally canceled, the chromatic $\beta$ perturbation propagates around the ring, giving rise to large nonlinear momentum dependencies of the betatron tune. A conventional design is to use local sextupoles generating a chromatic $\beta$ wave opposite to the one from each FFB, so they cancel each other out. A separate local correction is necessary on each side of the IP to avoid the chromatic beam smear at the IP. Studies are underway to use the chromatic interference of the two IRs to control chromatic beam smear around the ring and at both IPs. In case of a simultaneous operation of the two IRs, one can use the chromatic $\beta$ wave generated by one IR to cancel the chromatic kick of the other~\cite{marx}. This requires a certain betatron phase advance between the IRs. The phase advance can be adjusted in the arc section separating the two IRs. If the IRs are not identical the size of the $\beta$ wave can be adjusted using sextupoles between the IRs. The current progress is shown in Fig.~\ref{fig:chrom}.
\section{Summary}
The current design progress of the secondary IR for the EIC collider is presented in this paper. Some of the ongoing work includes continuing chromaticity correction, optimizing forward side magnets for an optimum neutron and proton acceptance, and evaluating the engineering feasibility on the IR magnets which has large apertures. Preliminary work has been already being done for the largest aperture quadrupole and the dipole in the downstream hadron beam-line.
%
%
%
\raggedend
\ifboolexpr{bool{jacowbiblatex}}%
	{\printbibliography}%
	{%
	

} 
%
%


\end{document}